\documentclass[aps,prx,twocolumn,unsortedaddress,floatfix]{revtex4}
\usepackage{amsthm}
\usepackage{amsfonts}
\usepackage{siunitx}
\usepackage{amsmath}
\usepackage{amssymb}
\usepackage{graphicx}
\usepackage{verbatim}
\usepackage[colorlinks]{hyperref}
\usepackage{tikz}
\usepackage{pgfplots}
\usepackage{braket}
\usepackage{xcolor}
\usepackage{physics}

\definecolor{linkcolor}{RGB}{0,83,166}
\hypersetup{
  colorlinks = true,
  allcolors = {linkcolor}
}

\begin{document}

 
\title{Programmable Exploration of Magnetic States in Lieb–Kagome Interpolated Lattices}
\author{Alejandro Lopez-Bezanilla$^1$}
\email[]{alejandrolb@gmail.com} 
\author{Pavel A. Dub$^2$} 
\author{Avadh Saxena$^1$}
\affiliation{$^1$Theoretical Division, Los Alamos National Laboratory, Los Alamos, New Mexico 87545, United States of America}
\affiliation{$^2$Chemistry Division, Los Alamos National Laboratory, Los Alamos, New Mexico 87545, United States of America}

\begin{abstract}
 We investigate a hybrid modeling framework in which a quantum annealer is used to simulate magnetic interactions in molecular qubit lattices inspired by experimentally realizable systems. Using phthalocyanine assemblies as a structurally constrained prototype, we model a continuous deformation from a Lieb to a kagome lattice, revealing frustration-driven disorder and magnetic field-induced reordering in the spin structure. The annealer provides access to observables such as the static structure factor and magnetization over a wide parameter space, enabling the characterization of magnetic arrangements beyond the reach of current molecular architectures. This surrogate modeling approach supports a feedback loop between experiment and programmable quantum hardware, offering a pathway to explore and iteratively design tunable magnetic states in synthetic quantum materials. The synthetic design, structural characterization, and quantum simulation framework established here defines a modular and scalable paradigm for probing the limits of engineered quantum matter across chemistry, condensed matter, and quantum information science.

\end{abstract} 

\date{\today}

\maketitle

\section{Introduction}

The search for controllable, scalable quantum systems has motivated extensive research into molecular qubits, whose intrinsic spin degrees of freedom and synthetic tunability render them compelling candidates for quantum information processing and quantum simulation~\cite{jenkins2016scalable,Kamlapure2018,PhysRevLett.114.106801}. 
An appealing candidate are phthalocyanines (Pc), chemically versatile, $\pi$-conjugated macrocycles that coordinate a wide range of metal ions, forming highly stable molecular complexes with well-defined spin states~\cite{Urdaniz2025}. When coordinated with open-shell transition metals, such as Mn, Fe, or Co~\cite{Suzuki2002}, they yield spin-bearing species that exhibit localized magnetic moments and tunable quantum properties. Among them, vanadyl phthalocyanine (VOPc) has emerged as a prototypical spin-1/2 molecular qubit, owing to its long coherence times, planar geometry, and compatibility with substrates~\cite{Malavolti2018,kavand2025}. Surface-deposited VOPc has been shown to self-assemble into square or quasi-square lattices on metallic substrates~\cite{Jin_2004,Blowey2019}, while preserving the integrity of the spin center and allowing for direct real-space characterization via STM spectroscopy~\cite{10.1063/5.0246931}.

Pc have been extensively studied in the context of molecular spintronics~\cite{PhysRevLett.105.077201,Suzuki2002}, Kondo physics~\cite{Ribeiro2024}, and single-molecule magnetism~\cite{GaitaArino2019}. Their modularity permits chemical tuning of ligand fields, spin multiplicity, and magnetic anisotropy~\cite{Coronado2020}, while their robust $\pi$-systems facilitate self-assembly into ordered two-dimensional arrays via van der Waals and $\pi$-$\pi$ interactions~\cite{Bonizzoni2017}. Although the typical packing geometry forms a full square lattice, engineered depletion—via selective desorption, tip manipulation, or patterned deposition—could yield Lieb-like geometries~\cite{Cui2020}, enabling exploration of flat-band magnetism and frustration-driven phenomena~\cite{Drost2017}. In such architectures, global interactions can be mediated by the substrate, including exchange coupling transmitted through the substrate’s electronic structure, as well as substrate-induced effects such as spin–orbit coupling or proximity-induced superconducting pairing~\cite{vanStraaten2018,Avvisati}.

\begin{figure}[htp]
\centering 
    \includegraphics[width=0.98\linewidth]{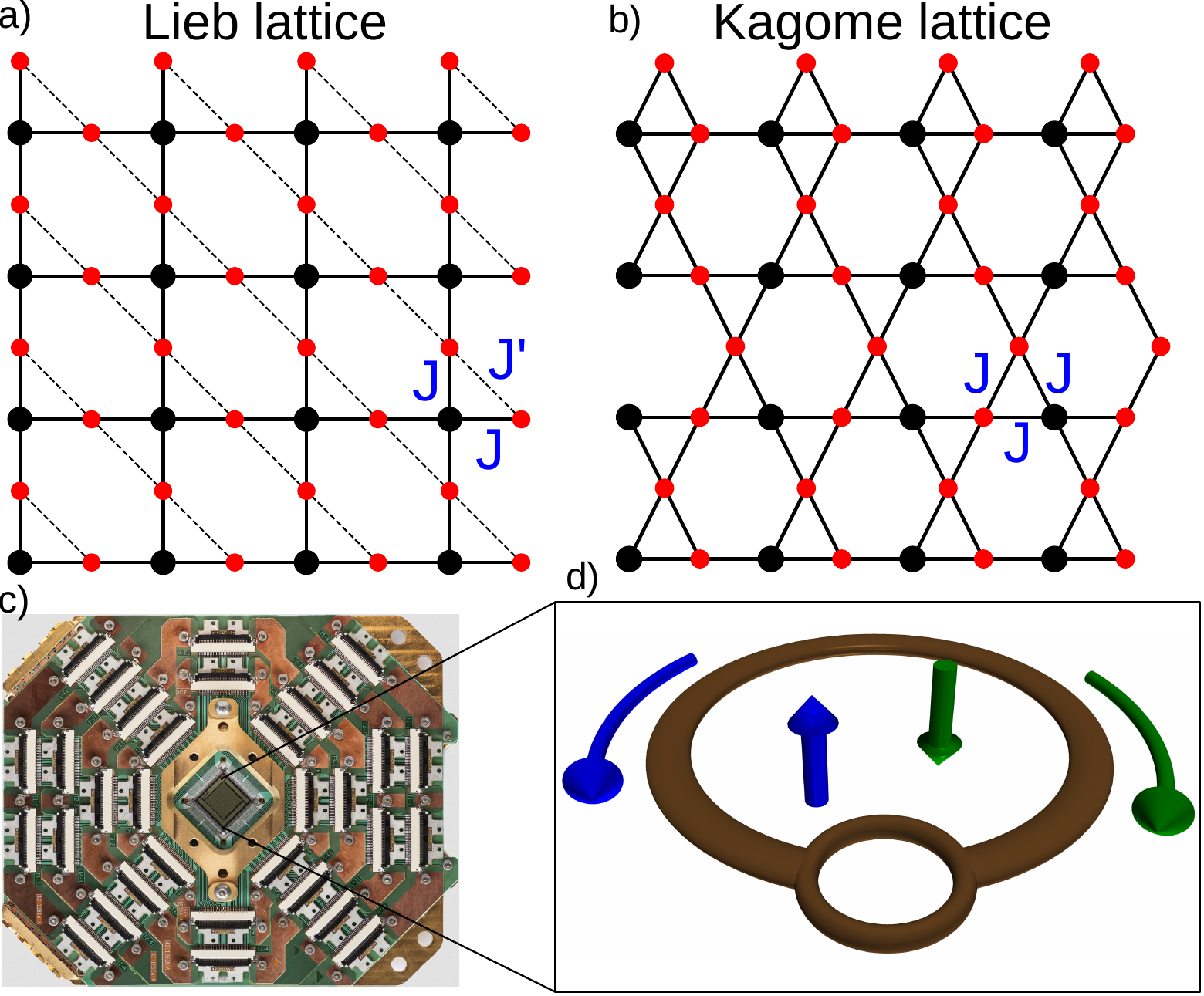}    
\caption{A Lieb lattice, (a), transforms into a kagome lattice, (b), through the application of horizontal shear strain, modeled here by a progressive increase in the coupling constant $J'$. 
The primary coupling $J$ (between red and black nodes) initially dominates over $J'$, until both become equal in the kagome geometry
(c) \textsc{D-Wave} Quantum annealing hardware used as a platform. (d) A schematic flux qubit showing opposite currents creating two different magnetic moments.
}
\label{fig:fig1}
\end{figure}

 \subsection{Lieb and Kagome Lattices} 
The Lieb lattice geometry is a nontrivial structure extensively studied in correlated electron systems—which differs from the full square lattice by a regular depletion of one out of every four sites, resulting in a unit cell with three magnetic centers. The Lieb lattice, composed of edge-centered and corner-sharing sites, supports a  non-frustrated magnetic bipartite structure that naturally hosts antiferromagnetic (AFM) order. Although it appears in two dimensions in some cuprate high-T$_c$-superconductors~\cite{Bednorz1986}, experimental realizations of these lattices have been attempted in various platforms, including ultracold atoms~\cite{Bercioux2009,Shen2010}, arrays of optical waveguides~\cite{Mukherjee2015}, surface patterning techniques~\cite{Slot2017}, and covalent-organic frameworks~\cite{phthalocyanine}, yet the ability to controllably deform one lattice into the other and monitor the corresponding evolution of magnetic order remains largely unexplored.

In contrast, the kagome lattice—formed by corner-sharing triangles—introduces intrinsic geometric frustration, precluding conventional magnetic ordering and giving rise to macroscopically degenerate ground states, spin liquids, and emergent gauge fields. Kagome physics has been experimentally realized in materials such as herbertsmithite ZnCu$_3$(OH)$_6$Cl$_2$~\cite{Khuntia2020}, a canonical quantum spin liquid candidate, and in kagome metals like FeSn ~\cite{Kang2020,PhysRevB.102.155103} and Co$_3$Sn$_2$S$_2$ ~\cite{Okamura2020}, which exhibit flat electronic bands, Dirac fermions, and anomalous Hall effects.

Despite the fact that Lieb and kagome lattices are topologically equivalent—each unit cell containing two edge-centered and one corner site—their magnetic ground states differ profoundly.  Importantly, these two lattice types can be connected through a continuous structural transformation that preserves topological equivalence while altering the degree of geometric frustration. While the Lieb lattice maintains a non-frustrated configuration through its bipartite connectivity, the kagome geometry induces frustration by introducing triangular loops of odd parity, which destabilize long-range order and give rise to a highly degenerate ground-state manifold. Studying the continuous deformation between these two geometries, such as through a shear-like distortion, provides a unique opportunity to isolate how subtle changes in lattice connectivity and symmetry control the onset of magnetic frustration, ground-state degeneracy, and external magnetic field responsiveness. This interpolation offers a tunable platform for probing the emergence of complex magnetic phenomena beyond what either geometry can support in isolation.

Exploring how the system transitions between these regimes—where frustration and order compete—requires control over lattice geometry and spin interactions that is difficult to achieve in molecular realizations. Although Pc assemblies provide a structurally coherent platform resembling a square lattice, they remain fundamentally rigid. 
To engineer a Pc array two strategies can be considered: (i) creating vacancies by selectively removing Pc molecules to eliminate specific spin sites~\cite{Khajetoorians2019}, or (ii) adding additional magnetic atoms or molecules at selected locations, such as the edge centers of square plaquettes, to effectively break translational symmetry and redefine the unit cell geometry. These modifications could be realized through controlled molecular desorption, atomic manipulation via scanning probe techniques, or targeted deposition using templated surfaces. Establishing such a geometry would enable the experimental realization of a Lieb-type magnetic lattice with tunable connectivity, suitable for probing magnetic frustration in a molecular setting.

\subsection{Utility of Quantum Annealer Platform}
A major concern is that the molecular arrangement is defined by steric packing, substrate interaction, and non-tunable intermolecular forces; once assembled, the geometry is fixed by thermodynamic and interfacial constraints. Any attempt to vary the lattice connectivity, coordination number, or interaction strength would require either new synthetic pathways or post-fabrication manipulation, while uncontrolled couplings make them ill-suited for exploring the continuum of Hamiltonians that define frustrated magnetic systems. This limitation motivates the use of an auxiliary modeling platform—such as a quantum annealer (QA)—which can reproduce and extend these geometries in a programmable, tunable, and experimentally tractable form.

In this context, QAs provide a powerful complementary platform—not as a replacement for experimental systems, but as a means of extending their capabilities into regimes that are difficult or impossible to access directly. Once experimental platforms have validated the underlying physical mechanisms—such as spin coherence, magnetic ordering, or lattice fidelity—the quantum annealer can be employed to complement and extend experimental capabilities, offering tunable control over lattice geometry, frustration, coupling asymmetry, and external fields. This enables systematic exploration of how subtle distortions or interaction imbalances influence the ground-state structure and spin correlations. In other words, the QA becomes a platform not merely for approximation or abstraction, but for deepening our understanding of the very physics that real materials hint at but cannot fully reveal. It allows us to disentangle structural and magnetic degrees of freedom, isolate specific interaction pathways, and probe transitions that are obscured in real materials by disorder, defects, or synthesis limitations.

In this work, we propose that QAs, with their programmable control over qubit connectivity and interaction strengths, offer a practical surrogate platform for modeling magnetic behavior in molecular lattices whose geometry cannot be readily modified post-synthesis. In particular, we focus on the continuous transformation from a Lieb lattice—experimentally approximated by modified Pc assemblies—to a kagome lattice, a transition that cannot be experimentally induced in such molecular systems due to their inherent structural rigidity. By tuning the spin-spin coupling asymmetry that mimics a shear deformation of the lattice (see Fig.~\ref{fig:fig1}), the quantum annealer enables us to simulate how spin correlations, magnetic order, and degeneracy evolve across this geometric pathway. This surrogate modeling approach allows us to anticipate how a Pc-based or similar magnetic material might behave if lattice connectivity could be continuously altered, and to identify which features—such as frustration thresholds or magnetic field-induced ordering—are robust against imperfections, and which depend critically on ideal geometry. As such, the QA does not merely reproduce existing molecular behavior, but extends it into a regime of counterfactual exploration, enabling targeted hypothesis generation for future synthetic strategies.
 
In the sense of providing a powerful avenue for extending QAs into regimes where experimental control is limited, quantum annealing occupies a role analogous to that of traditional numerical methods in materials science: just as density functional theory (DFT) and related computational techniques provided unprecedented access to the electronic, magnetic, and structural properties of materials from first principles,  QAs now offer an opportunity to model and interrogate many-body spin systems in a programmable, physically realized environment. But unlike purely numerical methods, QAs operate on quantum hardware and sample from physical, albeit engineered, quantum states—allowing them to act not only as solvers of effective models, but as experimental stand-ins for complex quantum matter. Their ability to emulate disorder, frustration, and geometry transitions in a fully tunable setting makes them uniquely suited to complement both {\it ab initio} theory and synthetic molecular systems, particularly in regimes where thermal fluctuations, degeneracies, and nonperturbative effects play dominant roles.

In this hybrid modeling framework, the QA becomes more than a simulator: it functions as a surrogate modeling tool in its own right, capable of validating theoretical models, guiding experimental design, and probing collective behavior in frustrated systems with high ground-state degeneracy—regimes where classical simulation techniques, such as Monte Carlo or exact diagonalization, may become computationally prohibitive. Unlike classical solvers, which often suffer from ergodicity breakdown or exponential scaling with system size and degeneracy, the QA samples directly from the physical low-energy manifold of an embedded Hamiltonian, leveraging quantum tunneling to explore configuration space more efficiently. This allows access to spin textures, disorder-induced transitions, and nontrivial correlations in parameter regimes that are either inaccessible or require uncontrolled approximations in classical treatments, thereby extending our reach into emergent magnetic behavior beyond what existing fabrication or simulation methods may be able to feasibly deliver.

Monte Carlo numerical techniques that are widely used to explore spin models and phase transitions, may suffer from slow convergence and sampling inefficiencies in highly frustrated or degenerate systems, where energy landscapes are rugged and the number of nearly degenerate configurations scales exponentially. In the energy minimization of frustrated spin systems, sampling inefficiencies often arise due to the so-called loop problem, where local update algorithms become trapped in metastable states and fail to efficiently access the global ground state. In such systems, closed loops of spins frequently contribute to frustration, and their relaxation requires coordinated multi-spin rearrangements. These collective updates are poorly captured by single-spin-flip dynamics, leading to slow convergence and incomplete sampling of the low-energy landscape. As a result, standard Monte Carlo methods frequently explore only a restricted subset of the configuration space, leading to premature convergence to metastable states. While techniques like simulated annealing or cluster updates can partially mitigate this issue, they scale poorly with system size and frustration complexity. By contrast, quantum annealing allows for coherent multi-spin tunneling across frustrated loops, providing more efficient access to low-energy states and enabling exploration of magnetic phases that are otherwise hidden to classical sampling methods.

QAs, by contrast, evolve quantum states through a transverse-field Ising model and naturally explore low-energy manifolds via quantum tunneling, which can provide enhanced sampling of classically hard-to-reach states. This is particularly advantageous in systems with macroscopic degeneracy, such as kagome-like spin liquids~\cite{Lopez-Bezanilla2023}. Moreover, QA provides direct access to physically realizable spin configurations under tunable Hamiltonians, offering a hardware-based approach to models that complement and extend purely numerical methods.

We use a QA as a tool enabling the exploration of magnetic and topological phases that are either experimentally inaccessible or structurally unstable in physical realizations. A QA by D-Wave Quantum Inc. is based on the programmable dynamics of spin‑1/2 systems in a transverse field Ising framework and provides a robust and reconfigurable environment in which to engineer on-demand lattices and precisely tune interactions, free from fabrication-induced variability. This analog computer permits us to probe the continuous geometry transformation between Lieb and kagome lattices—two geometrically distinct, yet topologically equivalent, frameworks relevant to flat-band physics, frustration, and emergent quantum order. Our findings extend the capabilities demonstrated by previous studies on superconducting qubit platforms~\cite{Kairys2020,king2021,Chamon2021}, further consolidating the QA’s role as an experimental proxy for exploring magnetic states.  While synthetic platforms can in principle host both lattice types, a controlled, real-space evolution between the two is challenging in practice. Here, we show that a QA not only replicates this transformation but also grants direct access to the low-energy landscape of associated magnetic states, establishing it as a powerful tool for testing hypotheses and guiding the design of molecular quantum devices. In doing so, we argue for a broader adoption of QA-based surrogate models as a practical and scalable extension of solid-state and molecular quantum science, capable of revealing quantum phases beyond the reach of laboratory synthesis and post-fabrication control.

In our model, the transformation is driven by a progressive tuning of the coupling parameter $J'$, which effectively mimics shear strain applied to the lattice. As $J'$ increases, the geometry transitions smoothly from an ordered Lieb configuration to a frustrated kagome-like structure. This is illustrated in Figure ~\ref{fig:fig1} that shows the schematic transformation from a Lieb lattice (left) to a kagome lattice (right), where nodes represent spin sites and their interactions. In the Lieb configuration, spins form a bipartite, non-frustrated geometry, where central nodes are initially twofold coordinated and connected to edge nodes via nearest-neighbor couplings. Diagonal couplings are initially set to zero. As the coupling parameter $J'$ increases, these diagonal terms gradually grow, effectively transforming the lattice geometry. The central nodes become fourfold coordinated, and the square symmetry is progressively distorted into a triangular framework, ultimately yielding a frustrated kagome lattice.

\section{Methods}

To simulate the magnetic behavior of geometrically evolving lattices, we encode a set of antiferromagnetically coupled spin-1/2 qubits onto the quantum processing unit (QPU) of a QA, using a time-dependent transverse-field Ising model of the form
\begin{equation}
\label{eq0}
\begin{aligned}  
    \mathcal{H}(s) = \mathcal{J}(s) \left( \sum_{i}h_{i}\sigma^z_i +\sum_{i,j}J_{ij}\sigma^z_i\sigma^z_j \right) - \Gamma(s) \sum_{i}\sigma^x_i  \,, 
\end{aligned}
\end{equation}
where $ \mathcal{J}$ and $\Gamma$ define the annealing schedule, $J_{ij}$ encodes the pairwise interaction between qubits $i$ and $j$, and $h_i$ represents a local longitudinal field applied on the $i^{th}$ node. 
The transverse field takes the system to a quantum superposition of states to subsequently collapse it into a classical state defined by a manifold of $\pm1$ variables. Here $\sigma^x$ and $\sigma^z$ are Pauli operators. 
The anneal process is controlled by the relative magnitude of classical $\mathcal{J}(s)$  and quantum $\Gamma(s)$ energy scale functions. During forward annealing and according to the evolution of the unitless annealing parameter $s$, $\Gamma(s=0)$ starts at its maximum while $\mathcal{J}$ vanishes. The process finishes when the transverse field is completely suppressed, $\Gamma(s=1) = 0$, and the state stops evolving.

\begin{figure}[htp]
\centering 
    \includegraphics[width=0.98\linewidth]{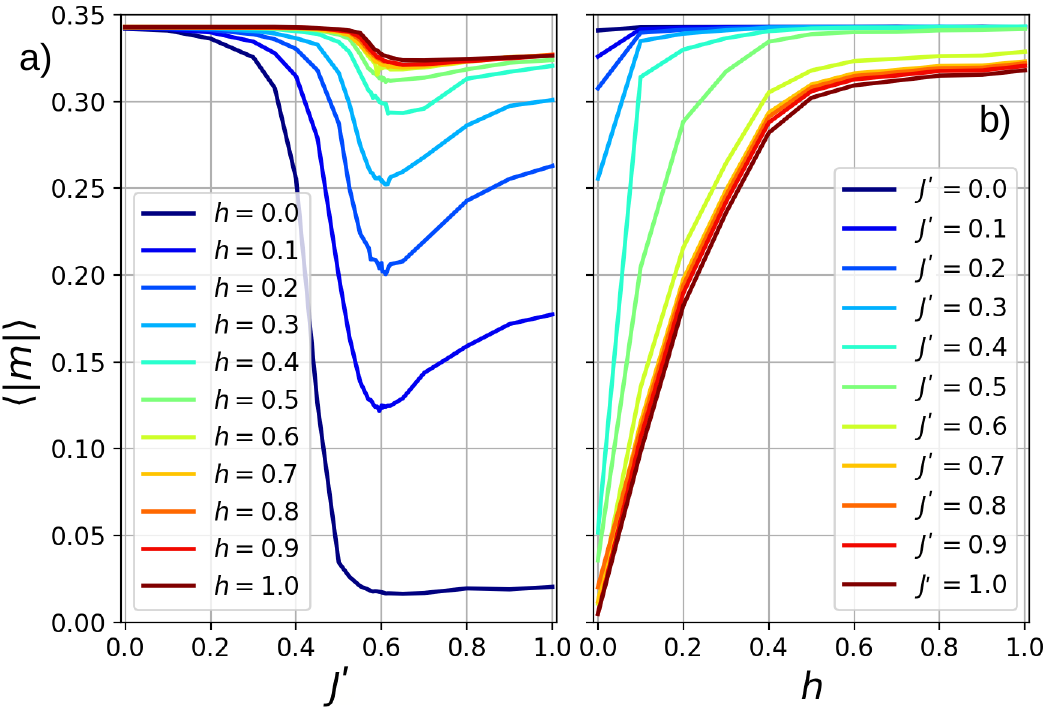}    
\caption{ (a) Evolution of the absolute value of average magnetization for different values of the local longitudinal field $h$ with $J'$ mimicking a finite Lieb lattice distortion. A value of $J=0.6$ defines the coupling between magnetic moments in horizontal and vertical lattice lines (see Fig.~\ref{fig:fig1}). Minimum of each $\langle|m|\rangle$ is found when $J'=J$. (b) shows the evolution of $\langle|m|\rangle$ as a function of the longitudinal field $h$ for fixed values of $J'$ coupling.
}
\label{fig:fig2}
\end{figure}

This formulation allows direct control over the strength and topology of spin couplings as well as the external symmetry-breaking terms. Through successive embeddings, we progressively transform a square Lieb lattice—composed of fourfold and twofold coordinated qubits—into a triangular kagome lattice characterized by fourfold coordination throughout (except for those nodes at the boundaries). This deformation is parameterized by tunable coupling sets of $J$ and $J'$ that govern the strength of the links exclusive to the Lieb lattice and act as a knob for inducing frustration. 
In contrast to real molecular systems where physical distortions result in unpredictable local fields and defect configurations, the annealer operates in a defect-free environment with precise qubit connectivity, offering a level of isolation and parameter fidelity unachievable in typical condensed matter systems. The resulting spin textures are interrogated via post-annealing readouts of qubit states, allowing us to reconstruct the real-space spin configurations, compute the average magnetization, and calculate static structure factors across a range of geometries and field conditions. 
This methodology provides a powerful route to directly visualize how topological equivalence can mask dramatically different magnetic behavior—a distinction that is crucial for both theoretical understanding and the practical design of programmable quantum materials.

Our system consists of 913 interconnected logical spins embedded on D-Wave's Advantage system. 
Each logical spin is constructed from three qubits coupled ferromagnetically with a coupling strength of $J_{FM}=-2$, enabling the embedding of higher-connectivity problem graphs onto the limited native connectivity of the quantum hardware. The total number of qubits used in this study is 2739. The reference coupling between constant-length triangle sides is set to $J=0.6$, while $J'$ ranges from $0.3$ to $1.7$. The Hamiltonian defining a classical state is:

\begin{equation}
\label{eq}
\begin{aligned}  
    H(s) =  \sum_{i,j}J'\sigma^z_i\sigma^z_j +  \sum_{i,k}J\sigma^z_i\sigma^z_k  \,, 
\end{aligned}
\end{equation}
where $i$ and $j$ are two neighboring spins (sitting at the red dots of Fig.~\ref{fig:fig1}) coupled by $J'$, and $i$ and $k$ are two neighboring spins coupled by $J$. The progressive deformation of one lattice into the other can be regarded as a continuous variation of $J'$ between a small value compared to $J$ (Lieb lattice) up to $J'=J$ (kagome lattice), and beyond ($J'>J$), where red-dot chains in Figure ~\ref{fig:fig1} become predominantly AFM.

To analyze spatial correlations and identify emergent symmetries, we compute the structure factor in momentum space as a post-processing tool applied to the spin configurations returned by the QA. The static structure factor, $S(\vec q)$, is defined by:
\begin{equation}
\label{eq2}
\begin{aligned}
    S(\vec q) = \frac{1}{N_s} \sum_{i,j} e^{-\vec{q}(\vec{r_i}-\vec{r_j})} \left\langle{\vec s}_i \cdot {\vec s}_j \right\rangle \,, 
\end{aligned}
\end{equation}  
where $N_s$ is the number of logical spins, $\vec{q} = (q_x,q_y)$ is a generic two-dimensional vector of the reciprocal space lattice, and $\vec{r_i}- \vec{r_j}$ is the vector between spins ${\vec s}_i$ and ${\vec s}_j$. Here $\left\langle \cdot \right\rangle$ denotes the ground state expectation value. $S(\vec q)$ helps to determine the structure of a magnetic material, including the directions in which moments point in an ordered spin arrangement, and the interaction between spins. The images we obtain is the magnetic structure analogue of the image that would be obtained by neutron diffraction techniques by reconstructions of the radiation scattered by the object ~\cite{lovesey1984}.

\section{Results}

Figure~\ref{fig:fig2}  shows how frustration, symmetry breaking, and external magnetic field shape the magnetic state of the system. The plot provides a quantitative analysis of the system’s magnetic response by plotting the average magnetization $\langle |m|\rangle$ as a function of the lattice deformation parameter $J'$, and as a function of the external longitudinal field $h$. Note that $J'$ controls the interpolation between a Lieb lattice ($J' \lessapprox 0.6$), a kagome lattice ($J'=0.6$), and beyond ($J'>0.6$), an oblique lattice. Each curve corresponds to a different fixed value of longitudinal field $h$. For $h=0$, the system begins in an antiferromagnetically ordered state in the Lieb geometry, where magnetization is partially suppressed due to perfect spin cancellation. Boundary spins add a magnetization above the nominal value of 0.33. As $J'$ increases toward 0.6, magnetic frustration is introduced, leading to a sharp drop in $\langle |m| \rangle$, consistent with the emergence of a disordered regime. This dip becomes shallower with increasing $h$, indicating that the external field lifts the degeneracy and restores partial magnetic ordering. At high field values ($h\geq0.6$), the magnetization remains nearly constant across all values of $J'$, suggesting that the field dominates over geometrical frustration, enforcing spin alignment regardless of underlying lattice symmetry.
This panel clearly delineates a frustration-driven suppression of order at intermediate $J'$ and low $h$, followed by field-induced magnetic stabilization, in excellent agreement with the structure factor evolution shown later.

\begin{figure*}[htp]
\centering
    \includegraphics[width=0.98\textwidth]{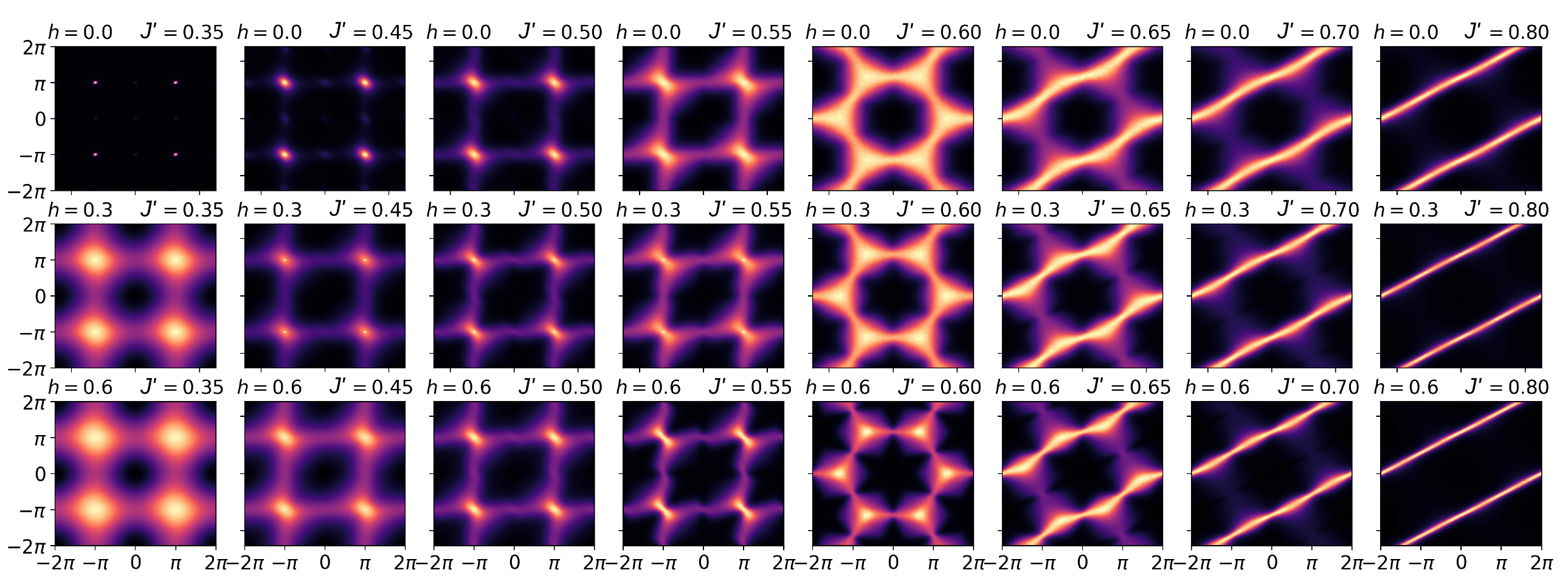}
\caption{ Static structure factor $S(\vec q)$ plotted for a range of coupling ratios $J'$ and longitudinal fields $h$, illustrating the magnetic evolution of a spin system embedded in a programmable lattice. The first four columns are plotted using the Brillouin zone of the square Lieb lattice, while the last four use the reciprocal lattice of the hexagonal (kagome) geometry.  At low $J'$ and zero field, the system exhibits long-range AFM order with Bragg peaks at $(\pi,\pi)$, characteristic of the crystalline Lieb lattice. Increasing $J'$ to 0.6 suppresses this order, leading to a disordered, highly frustrated regime with diffuse correlations—indicative of a spin liquid–like state. Upon increasing $h$ while keeping $J'$ fixed, magnetic order is restored via magnetic field-induced crystallization, selectively favoring a subset of configurations from the degenerate manifold. Further increasing $J'$ at constant field drives the system into a new ordered state, now governed by dominant diagonal couplings.  
}
\label{fig:fig3}
\end{figure*}   

Figure~\ref{fig:fig2}b provides a complementary perspective, showing how magnetization, $\langle|m|\rangle$, evolves as a function of the longitudinal field $h$ for fixed values of $J'$. For small $J'=0.35$, corresponding to a Lieb-like geometry with weak diagonal couplings ($J'<J$), the system begins in a robust AFM state. The magnetization increases slowly with $h$, reflecting the resistance of a non-frustrated lattice to magnetic field-driven spin polarization. As $J'$ increases toward $J'=0.6$, where the diagonal and edge couplings become equal ($J=J'$) and the lattice geometry reaches the perfect kagome configuration, the system enters a maximally frustrated regime. Here, magnetization responds more rapidly to small values of $h$, indicating that the degeneracy of the ground-state manifold makes the system highly susceptible to symmetry-breaking fields. 

Further increasing $J'$ introduces a distortion of the kagome lattice geometry that stabilizes a different form of AFM order, and the magnetization curves begin to converge and saturate with increasing $h$. This behavior underscores the significance of the $J'=J$ point as a critical crossover in both geometry and magnetism, marking the regime of maximal frustration and strongest field sensitivity. The results reveal how the interplay between lattice connectivity and external field drives a transition from a stable antiferromagnet to a disordered, field-responsive phase. Importantly, the quantum annealer captures this tunability with precision, providing a model for how real materials could behave under controlled geometric distortions or applied fields.

Similar to conventional condensed matter techniques—such as neutron scattering or resonant X-ray scattering—a QA can produce observables but with vastly greater flexibility and efficiency. In particular, the static structure factor $S(\vec q)$, which encodes information about spin correlations and magnetic order in momentum space, can be extracted directly from the post-annealing qubit configurations for any set of model parameters. 
Unlike physical experiments that require sophisticated instrumentation, long measurement times, and carefully engineered samples, the QA allows us to compute $S(\vec q)$ for an arbitrary number of configurations, geometries, and fields with minimal effort. This opens the door to a systematic, high-resolution exploration of the magnetic phase diagram across a continuous parameter space. In what follows, we present structure factor maps for a range of $J'$ and longitudinal field $h$, revealing the progression from long-range AFM order in the Lieb lattice to frustrated, disordered regimes in the kagome limit.

Figure ~\ref{fig:fig3} presents a systematic exploration of the magnetic structure factor  $S(\vec q)$, where the $\vec q$ dependence of the intensity provides information on the ground-state spin correlations across a landscape defined by two key parameters: the relative diagonal coupling $J'$ and the external longitudinal field $h$. 
The figure consists of eight columns representing different coupling ratios, with each row corresponding to an increasing value of the applied longitudinal field. From left to right, the first four columns are plotted using the reciprocal lattice of the square Lieb geometry, reflecting the square lattice symmetry in the weak-diagonal-coupling regime. In contrast, the rightmost four columns are computed using the reciprocal lattice vectors of the hexagonal (kagome) lattice, appropriate for the strong-frustration regime where the underlying direct lattice approximates triangular coordination. This shift in Brillouin zone representation enables accurate comparison of momentum-resolved spin correlations within the respective geometric frameworks, enabling us to trace a multi-phase evolution of the spin system as it transitions through distinct magnetic regimes.

At $h=0$ and a small $J'=0.35$, the system exhibits a clean, long-range AFM order. This is reflected in the static structure factor which displays sharp, well-defined Bragg peaks centered at high-symmetry points such as $(\pi,\pi)$, indicating strong spin–spin correlations and a unique ground state (up to time-reversal symmetry breaking) stabilized by the bipartite lattice geometry. In this regime, the diagonal coupling $J'$ is significantly weaker than the principal coupling $J$. As a result, the magnetic interactions are effectively governed by the square lattice connectivity, and the weak diagonal terms do not introduce significant frustration or alter the ground-state ordering.

As the coupling  $J'$ increases to 0.6 while keeping $h=0$, the system crosses with no phase transition into a regime of maximal magnetic disorder. The formerly crystalline structure dissolves into a diffuse $S(\vec q)$ profile, indicative of a massively degenerate ground-state manifold. This regime corresponds to the kagome lattice, with local frustration destroying global ordering. It is here that the system approximates a classical spin liquid~\cite{ANDERSON1196} or disordered magnetic plasma, consistent with theoretical descriptions of frustrated Ising systems~\cite{Lopez-Bezanilla2023}.

Holding $J'=0.6$ fixed and gradually increasing the longitudinal field $h$ from 0 to 0.3 and 0.6 induces a re-emergence of magnetic order. The application of $h$ breaks time-reversal symmetry and selects among previously degenerate configurations, energetically favoring specific spin arrangements. This symmetry-breaking field effectively crystallizes the disordered state into a magnetically ordered phase, as evidenced by the reappearance of well-defined intensity peaks in $ S(\vec q)$. This crystallization is analogous to magnetic field-induced ordering observed in artificial spin ice and transverse-field Ising models~\cite{RevModPhys.85.1473,Lopez-Bezanilla2023,Lopez-Bezanilla2024}. 
With the magnetic field fixed at $h=0.6$, further increasing $J'$ leads to another transformation in the spin texture. 

As $J'$  increases beyond 0.6, the effective geometry departs from the symmetric kagome limit, where the equilateral triangle evolves into an asymmetrically scalene ($J'>0.6$), in which the balance between edge and diagonal interactions is broken. This distortion lifts residual degeneracies and favors new spin alignment patterns, effectively selecting a different AFM ground state stabilized by a dominant diagonal $J'$ coupling that controls the strength and spatial extent of spin correlations. This is a different but still structured ordering pattern distinct from the initial Lieb limit but no longer disordered. The result is a third magnetic crystal, shaped not by field selection but by lattice-driven interaction hierarchy, as observed in the last column of Figure~\ref{fig:fig3}.

This multi-phase evolution reveals the power of combining geometrical and field-based controls within a quantum annealing framework. It shows how a programmable quantum system can emulate not only ground states but also transitions between fundamentally different magnetic regimes—ranging from highly ordered to maximally frustrated, and back into re-crystallized or field-pinned configurations. In doing so, it provides insight into the subtle interplay between lattice symmetry, frustration, and external perturbations—offering an experimental venue to probe possible phase transitions that are difficult to realize or resolve in actual materials. Notably, the fact that both field and geometry can independently drive or stabilize order opens a design space where synthetic spin models can be crafted to explore nontrivial emergent behavior and test theoretical models of frustration-induced magnetism under controlled conditions.

\section{Conclusions}
\label{sec:sum} 

We have proposed and implemented a hybrid approach in which a QA is used to model interaction patterns and magnetic responses that could be observed in molecular qubit systems that are experimentally realizable but structurally rigid. Using a realistic Lieb-like geometry as the point of departure, we emulate a continuous deformation toward frustrated kagome configurations and investigate the evolution of spin correlations and field response across this geometric manifold. The static structure factor and magnetization profiles extracted from the annealer reveal smooth configuration transitions driven by frustration and symmetry breaking, including disorder-induced suppression of magnetic order and field-stabilized crystallization of degenerate spin configurations.

This environment allows us to emulate spin systems with continuously tunable geometry and interaction hierarchy, using a transverse-field Ising model embedded in quantum hardware. The annealer mimics the relevant physics of the molecular platform in a more accessible and adaptable form, enabling us to simulate frustrated configurations, drive symmetry breaking, and probe emergent order in parameter regimes that extend beyond current experimental reach.
The ability to systematically tune geometry and external fields in a controlled environment positions quantum annealing as a viable modeling platform for exploring quantum magnetic phases inaccessible in current synthetic systems. Beyond acting as a simulator, the annealer enables rapid prototyping of spin Hamiltonians under physical constraints, with observables comparable to those obtained in large-scale scattering experiments. This framework supports a feedback-driven design strategy, where experimentally measurable parameters constrain simulation, and simulation results inform chemical or structural optimization. Such a loop offers a path toward programmable quantum matter, in which computational and synthetic layers are co-developed to access emergent regimes beyond current material realizations.
 
We believe that the QA thus serves not merely as a numerical solver, but as a surrogate quantum model—playing a role analogous to that of density functional theory (DFT) in electronic structure, but tailored to interacting spin systems and emergent magnetism. Unlike traditional computational techniques, the QA samples from physically instantiated quantum states, yielding direct access to observables such as magnetization, which are normally reserved for large-scale scattering experiments. Here, we demonstrate that these quantities can be systematically computed across a continuous landscape of geometries and fields, making the QA an efficient and versatile tool for exploring phase transitions, degeneracy lifting, and magnetic field-induced ordering in synthetic spin networks.

A key novelty of our approach lies in the possibility of establishing a closed-loop design strategy between experimental realization and quantum simulation. Parameters measured from the molecular system—such as coherence times, coupling asymmetries, or noise thresholds—could inform and constrain the Hamiltonians embedded into the QA, ensuring the physical relevance of the simulated models. In return, QA simulations can identify magnetic regimes that are both theoretically interesting and experimentally accessible, suggesting how molecular design could be refined to stabilize or access specific phases. This feedback mechanism transforms the QA from a passive simulator into an active guide for materials development, shaping hypotheses, prioritizing synthesis targets, and enabling iterative refinement.

Looking ahead, this hybrid methodology offers a scalable and modular path toward programmable quantum matter. As quantum annealing hardware evolves—offering higher connectivity, lower noise, and more expressive Hamiltonians—its modeling capacity will continue to expand. Simultaneously, advances in synthetic chemistry and atomic-scale fabrication may unlock new ways to introduce tunability into molecular lattices themselves. In this vision,  QAs do not merely model what already exists; they help chart the landscape of what could be made next, bringing us closer to a design framework in which molecular synthesis and quantum simulation co-evolve in pursuit of functional quantum materials.
 
\section{Acknowledgments}
ALB is grateful to Scott Crooker for engaging discussions and thoughtful input.
Research presented in this article was supported by the Laboratory Directed Research and Development (LDRD) program of Los Alamos National Laboratory (LANL) under project number 20200056DR and 20240343ER.
Los Alamos National Laboratory is managed by Triad National Security, LLC, for the National Nuclear Security Administration of the U.S. Department of Energy under Contract No. 89233218CNA000001.

\bibliographystyle{apsrev4-1}

 \end{document}